\documentclass[prc,preprint,showpacs,preprintnumbers,amsmath,amssymb]{revtex4}

\usepackage[mathscr]{eucal}
\usepackage{graphicx}
\def\slr#1{\setbox0=\hbox{$#1$}           
   \dimen0=\wd0                                 
   \setbox1=\hbox{/} \dimen1=\wd1               
   \ifdim\dimen0>\dimen1                        
      \rlap{\hbox to \dimen0{\hfil/\hfil}}      
      #1                                        
   \else                                        
      \rlap{\hbox to \dimen1{\hfil$#1$\hfil}}   
      /                                         
   \fi}

\def\be{\begin{eqnarray}}
\def\ee{\end{eqnarray}}

\renewcommand{\theequation}%
    {\arabic{section}.\arabic{equation}}
\makeatletter \@addtoreset{equation}{section} \makeatother

\begin{document}

\preprint{BCCNT: 04/091/327}

\title{Quark Propagation in the Presence of a $<A_\mu^aA_a^\mu>$ Condensate}

\author{Xiangdong Li}
\affiliation{%
Department of Computer System Technology\\
New York City College of Technology of the City University of New
York\\
Brooklyn, New York 11201 }%

\author{C. M. Shakin}
\email[email address:]{casbc@cunyvm.cuny.edu}

\affiliation{%
Department of Physics and Center for Nuclear Theory\\
Brooklyn College of the City University of New York\\
Brooklyn, New York 11210
}%

\date{Sept, 2004}

\begin{abstract}
There is a good deal of current interest in the condensate
$<g^2A_\mu^aA_a^\mu>$ which has recently be shown to be the Landau
gauge version of a more general gauge-invariant expression. In the
present work we consider quark propagation in the presence of such
a condensate which we assume to be present in the vacuum. We
describe the vacuum as a random medium of gluon fields. We discuss
quark propagation in that medium and show that the quark
propagator has no on-mass-shell pole indicating that a quark
cannot propagate over extended distances. That is, the quark is a
nonpropagating mode in the gluon condensate.

\end{abstract}


\maketitle

\section {Introduction}

Some years ago we presented a discussion of the properties of the
gluon condensate [1,2]. Central to our work was the specification
of the condensate $<g^2A^2>$ which was obtained from the known
value of
$<0\mid(\alpha_s/\pi)G_a^{\mu\gamma}G^a_{\mu\gamma}\mid0>$. (The
value of the latter quantity was the only parameter we needed for
our analysis.) We obtained a dynamical gluon mass of $m_G$=649 MeV
and a dynamical quark mass of $m_q^{Gl}$=432 MeV, which arose from
coupling to the gluon condensate [1,2]. In addition, we predicted
a glueball mass of $1.0\pm0.15$ GeV.

Our work was limited by the fact that the condensate $<g^2A^2>$
was thought to be non gauge invariant. However, recent years have
shown a renewed interest in such a condensate. For example, we
note that the operator $G_{\mu\nu}^aG_a^{\mu\nu}$ generates terms
of order $1/p^4$ in the operator product expansion, while $A_\mu
A^\mu$ is the only term that has a vacuum expectation value which
will generate $1/p^2$ terms in studies making use of the operator
product expansion. There is significant empirical evidence for the
importance of such $1/p^2$ terms [3].

The issue of gauge invariance has been discussed by several
authors [4,5] and it was argued that $<A^2_\mu>$ may be important
for the study of the topological structure of the vacuum and quark
confinement [4,6,7]. Kondo [4] was responsible for introducing a
BRST-invariant condensate of dimension 2 \be
\mathcal{O}=\frac{1}{\Omega}<\int
d^4x\,\mbox{Tr}\left(\frac{1}{2}A_\mu(x)\cdot A_\mu(x)-\alpha i
\,c(x)\cdot \bar{c}(x)\right)> \ee where $c(x)$ are Faddeev-Popov
ghosts, $\alpha$ is the gauge-fixing parameter and $\Omega$ is the
integration volume. To quote Kondo: ``It is clear that
$\mathcal{O}$ reduces to $A^2_{min}$ in the Landau gauge
$\alpha=0$. Therefore, the Landau gauge turns out to be a rather
special case in which we do not need to consider the ghost
condensation." Here, the minimum value of the integrated squared
potential is $A^2_{min}$, which has a definite physical meaning
[4].

Recently Arriola, Bowman and Broniowski [8] have made use of
lattice data for the quark propagator. They have considered an
operator product expansion (OPE) in the analysis of the QCD
lattice data, including $<A^2>/Q^2$ terms and use the relation
$m_G^2=(3/32)g^2<A^2>$ to obtain $m_G=(625\pm 33)$ MeV, a
numerical result that is consistent with the one we have obtained
by a different method [1,2].

Our work contains the following material. In Section II we will
introduce a description of the QCD vacuum as a random medium and
define our model for the matrix elements of the condensate field.
In Section III we review the theory of wave propagation in a
random medium and introduce the quark propagator. In Section IV we
calculate the quark self-energy, $\Sigma(p)=A(p^2)\slr p +B(p^2)$,
and present results for $M(p^2)=(B(p^2)+m_q^{cur})/(1-A(p^2))$,
$B(p^2)+m_q^{cur}$ and $[1-A(p^2)]$ in Figs. 2-14. Section V
contains some further discussion and conclusions.

\section {The QCD Vacuum as a Random Medium}

We now proceed to consider quark propagation in a random medium
which is characterized, in part, by finite value of the $<A^2>$
condensate. It is useful to separate the vector potential into a
condensate field and a fluctuating field. We write \be A_i^a(x) =
\mathbb{A}^a_i(x)+\mathcal{A}_i^a(x).\ee This division of the
vector potential into a condensate field, which we assume may be
treated as a classical random field, and a fluctuation field,
$\mathcal{A}_\mu^a(x)$, will be useful. The basic idea is that,
independent of the nature of the condensate, be it a monopole
vortex or instanton condensate, there is a natural separation of
the field into a nonperturbative condensate field and a
fluctuation field which can be treated perturbatively. (Some
discussion of this separation is given in Shuryak's book with
reference to its use in a QCD sum-rule analysis [9].)

We introduce a vacuum state which has the following properties:
\be <vac\mid\mathbb{A}_a^i(0)\mid vac>=0, \ee \be
<vac\mid\mathbb{A}_a^i(0)\mathbb{A}_b^j(0)\mid vac> =
\frac{\delta_{ij}}{3}\phi_0^2\frac{\delta_{ab}}{8},\ee etc.

We will find it useful to generalize this result to provide a
covariant description of the vacuum state. We define a vacuum
state, $\widetilde{\mid vac>}$, which has the following properties
\be
<\widetilde{vac}\mid\mathbb{A}_\mu^a(0)\mid\widetilde{vac}>=0,\ee
\be <\widetilde{vac}\mid
g^2\mathbb{A}_\mu^a(0)\mathbb{A}^b_\nu(0)\mid\widetilde{vac}>
=-\frac{g_{\mu\nu}}{4}g^2\phi_0^2\frac{\delta_{ab}}{8},\ee \be
<\widetilde{vac}\mid\mathbb{A}_\mu^a(0)\mathbb{A}_\nu^b(0)\mathbb{A}^c_\rho(0)\mid\widetilde{vac}>=0,\ee
\be  <\widetilde{vac}\mid
g^4\mathbb{A}_\mu^a(0)\mathbb{A}_\nu^b(0)\mathbb{A}^c_\rho(0)\mathbb{A}^d_\eta(0)\mid\widetilde{vac}>\,\,\,\,\,\,\,\,\,\,\,\,\,\,\,
\,\,\,\,\,\,\,\,\,\,\,\,\,\,\,\,\,\,\,\,\,\,\,\,\,\,
\\\nonumber =\,
<\widetilde{vac}\mid\mathbb{A}_\mu^a(0)\mathbb{A}^b_\nu(0)\mid\widetilde{vac}><\widetilde{vac}\mid
\mathbb{A}_\rho^c(0)\mathbb{A}^d_\eta(0)\mid\widetilde{vac}>
\\\nonumber + <\widetilde{vac}\mid
\mathbb{A}_\mu^a(0)\mathbb{A}^c_\rho(0)\mid\widetilde{vac}><\widetilde{vac}\mid
\mathbb{A}_\nu^b(0)\mathbb{A}^d_\eta(0)\mid\widetilde{vac}>
\\\nonumber+<\widetilde{vac}\mid
\mathbb{A}_\mu^a(0)\mathbb{A}^d_\eta(0)\mid\widetilde{vac}><\widetilde{vac}\mid
\mathbb{A}_\nu^b(0)\mathbb{A}^c_\rho(0)\mid\widetilde{vac}>
 \\ = (g_{\mu\nu}g_{\rho\eta}\delta_{ab}\delta_{cd}
 +g_{\mu\rho}g_{\nu\eta}\delta_{ac}\delta_{bd}+g_{\mu\eta}g_{\nu\rho}\delta_{ad}\delta_{bc})g^4\phi_0^4/(16\cdot
64), \ee etc. Here, all odd correlation functions are taken to be
zero  and all even correlation functions may be expressed in terms
of the two-point correlation function, as in Eq. (2.8). This
characterization is that of a Gaussian random medium.

More generally, we might put \be <\widetilde{vac}\mid
g^2\mathbb{A}_\mu^a(x)\mathbb{A}^b_\nu(y)\mid\widetilde{vac}>
=-\frac{g_{\mu\nu}}{4}g^2\phi_0^2\frac{\delta_{ab}}{8}\exp
\left[-\mid(x-y)^2\mid ^{1/2}/L \right] \ee where $L$ is a
correlation length. The model we will study has $L \rightarrow
\infty$. That is, $L$ will be taken to be large compared to the
characteristic mean-free-path associated with the damping of the
quark propagator.

We wish to study the equation for the quark field \be
(i\gamma^\mu\partial_\mu-g\mathbb{A}^\mu_a(x)\gamma_\mu\frac{\lambda^a}{2})\psi(x)=0\ee
and the associated quark propagator. We have only included the
condensate field in Eq. (2.10) and will treat
$\mathbb{A}_\mu^a(x)$ as a random field having the properties
described above. With that goal in mind, we will review the
classical theory of wave propagation in random media.

\section {Wave Propagation in a Random Medium}

A useful review of wave propagation in random media has been given
by Frisch [10]. We follow the notation of that work, with a few
minor exceptions. One of the more elementary problems, which is
characteristic of the statistical analysis, is that of scalar wave
propagation in a time-independent, lossless, homogeneous,
isotropic random medium. The field satisfies the Helmholtz
equation \be \nabla^2 \varphi(\vec{r})
+k_0^2n^2(\vec{r})\varphi(\vec{r})=\delta(\vec{r}),\ee where
$n^2(\vec{r})$ is the index of refraction and $k_0$ is the
free-space wave number. One writes [10] \be
n^2(\vec{r})=1+\mu(\vec{r}),\ee where $\mu(r)$ is a centered
homogeneous and isotropic random function with finite moments.
That is $<\mu(\vec{r})>=0$, while the correlation function \be
\Gamma(\mid \vec{r}-\vec{r'}\mid)=<\mu(\vec{r})\mu(\vec{r'})> \ee
is unequal to zero. Here the brackets denote an ensemble average.
For simplicity, we can assume that $\mu(\vec{r})$ is a Gaussian
random function.

The Green's function satisfies a random integral equation. With
\be G^{(0)}(\vec{r},\vec{r'})\equiv
-\frac{\exp{\{ik_0\mid\vec{r}-\vec{r'}\mid}\}}{4\pi\mid\vec{r}-\vec{r'}\mid}\ee
we have \be
G(\vec{r},\vec{r'})=G^{(0)}(\vec{r},\vec{r'})-k_0^2\int
G^{(0)}(\vec{r},\vec{r_1})\mu(\vec{r_1})G(\vec{r_1},\vec{r'})d\vec{r_1}.\ee

We write the last equation as \be G = G^{(0)}-G^{(0)}L_1G,\ee
where $L_1$ is the random element. We are interested in the
ensemble average of $G$, which we denote as $<G>$. Thus \be
<G>=G^{(0)} - <G^{(0)}L_1G> ,\ee since  $<G^{(0)}> = G^{(0)}$.
(Here $k_0^2$ has been absorbed in $L_1$.)

One may develop a diagrammatic analysis of a perturbative
expansion of Eq. (3.7) [10]. When averaging the various terms in
such an expansion over the ensemble one finds it useful to
introduce so-called p-point correlation functions
$\Gamma_p(1,2,\cdot\cdot\cdotp)$: \be\Gamma_1(1) & =& <\mu(1)>,
\\ &=& 0,\ee
\be \Gamma_2(\vec{1},\vec{2})=<\mu(\vec{1})\mu(\vec{2})>,\ee \be
\Gamma_3(\vec{1},\vec{2},\vec{3})=<\mu(\vec{1})\mu(\vec{2})\mu(\vec{3})>,\ee
\be
\Gamma_4(\vec{1},\vec{2},\vec{3},\vec{4})=<\mu(\vec{1})\mu(\vec{2})\mu(\vec{3})\mu(\vec{4})>
-\Gamma_2(\vec{1},\vec{2})\Gamma_2(\vec{3},\vec{4})\\\nonumber
-\Gamma_2(\vec{1},\vec{4})\Gamma_2(\vec{2},\vec{3})-
\Gamma_2(\vec{1},\vec{3})\Gamma_2(\vec{2},\vec{4}),\ee
 etc. For a Gaussian centered random function
$\Gamma_1=0, \Gamma_3=0, \Gamma_5=0$, etc.

A simple first-order smoothing approximation is \be
<G(\vec{r},\vec{r'})>=G^{(0)}(\vec{r},\vec{r'})+k_0^4\int
G^{(0)}(\vec{r},\vec{1})G^{(0)}(\vec{1},\vec{2})\Gamma(\vec{1},\vec{2})<G(\vec{2},\vec{r'})>d\vec{1}
d\vec{2},\ee in the notation of Ref. [10]. (See Eq.(4.33) of
[10].) This approximation is most appropriate for a small
(generalized) Reynolds number defined as \be R=\epsilon k_0^2
L,\ee where $\epsilon$ is some (dimensionless) measure of the
scale of $\mu$ $[\mu=\epsilon\mu^{*}$, where $\mu^{*}=O(1)]$.

We will here be interested in the case of large generalized
Reynolds number. In this case Eq. (3.13) is replaced by the
Kraichnan equation [11] \be <G(\vec{r},\vec{r'}> =
G^{(0)}(\vec{r},\vec{r'})+k_0^4\int
G^{(0)}(\vec{r},\vec{1})<G(\vec{1},\vec{2}><G(\vec{2},\vec{r'})>\Gamma(\vec{1},\vec{2})d\vec{1}
d\vec{2},\ee or in a more abstract notation \be
<G>=G^{(0)}+G^{(0)}<L_1<G>L_1><G>.\ee (See Eqs.(4.76) and (4.79)
of Ref. [10].) Kraichnan [11] has shown that Eq. (3.7) may be
reduced to Eq. (3.16) upon the introduction of an additional
stochastic element (random coupling between different Fourier
components of the field). It has also been shown [11] that Eq.
(3.16) will give sensible results, since it can be considered both
as the solution of a model equation and as approximate solution of
Eq. (3.7).

It is useful to introduce a self-energy operator, $\Sigma$. From
Eq. (3.15), we have \be <G(\vec{r},\vec{r'})> =
G^{(0)}(\vec{r},\vec{r'})+k_0^2\int
G^{(0)}(\vec{r},\vec{1})\Sigma(\vec{1},\vec{2})<G(\vec{2},\vec{r'})>d\vec{1}
d\vec{2},\ee where \be
\Sigma(\vec{1},\vec{2})=<G(\vec{1},\vec{2})>\Gamma(\vec{1},\vec{2}).\ee
As usual, this equation is particularly simple in momentum space,
where we can write \be <G(p)> =
\frac{1}{[G^{(0)}(p)]^{-1}-\Sigma(p)},\ee in the case of an
infinite medium. The equations for the propagator and for
$\Sigma(p)$ are shown in Fig. 1. There we also show the kind of
diagrams which are being summed in the random-coupling
approximation.

\begin{figure}
\includegraphics[bb=0 100 280 400, angle=0, scale=1]{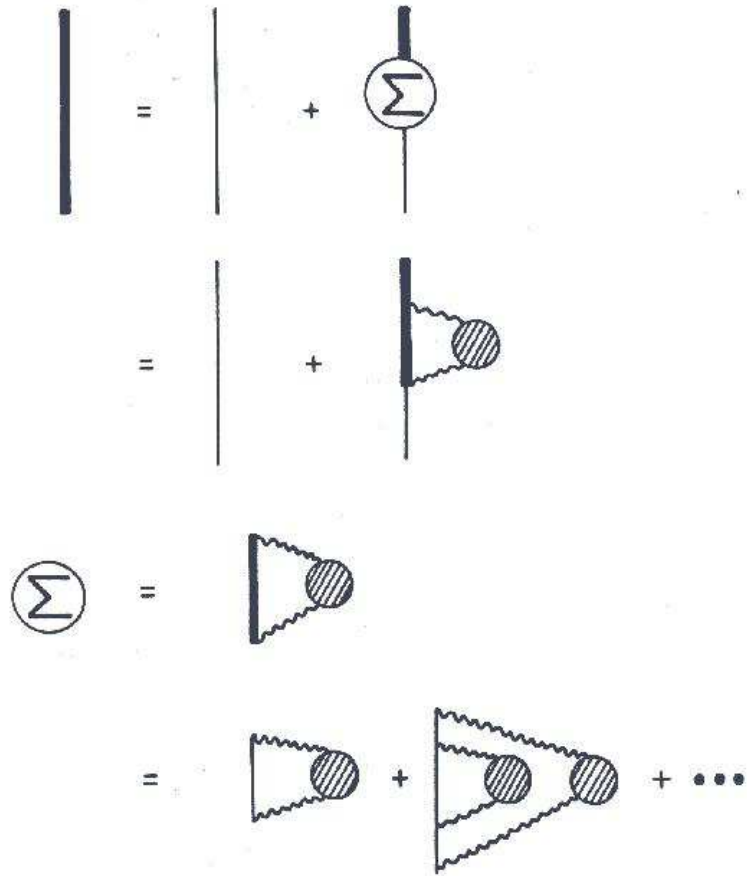}%
\caption{Equations for the ensemble-averaged quark propagator and
the self-energy in the random-coupling model. a) The solid line is
the full propagator and the light line is the propagator $[\slr
p-m_q^{cur}]^{-1}$. b) A self-consistent model for the
(irreducible) self-energy $\Sigma(p^2)$. The cross-hatched circle
denotes the condensate and the wavy lines are condensate gluons of
zero momentum. c) Here we show the type of nested diagrams which
are summed in this model for the self-energy $\Sigma(p^2)$.}
\end{figure}

The translation of this analysis to a study of the quark
propagator is immediate. We define a free propagator,
$S^{(0)}(x,x')$, and the ensemble average of the correlation
function of the Heisenberg fields: \be
<S(x,x')>=<\widetilde{vac}\mid
T(\psi(x)\bar{\psi}(x'))\mid\widetilde{vac}>.\ee The random
element, $L_1$, is \be
L_1=g\mathbb{A}_a^\mu(x)\gamma_\mu\lambda^a/2.\ee We write \be
\Gamma_{\mu\nu}^{ab}(1,2) &=& <\widetilde{vac}\mid
g^2\mathbb{A}_\mu^a(0)\mathbb{A}_\gamma^b(0)\mid\widetilde{vac}>
\\&=& -\frac{g_{\mu\nu}}{4}g^2\phi_0^2\frac{\delta_{ab}}{8}.\ee
Here we have generalized the scalar correlation function,
$\Gamma(1,2)$, to a form appropriate in the case the correlation
functions involves random functions with color and Lorentz
indices. In momentum space we have \be S^{(0)}(p) = \frac{1}{\slr
p},\ee and \be <S(p)>\,=\frac{1}{\slr p-\Sigma(p)}.\ee

We need not include an $i\epsilon$ in the denominators on the
right-hand side of Eqs. (3.24) and (3.25) since we will consider
solutions for $<S(p)>$ which have no singularities for real $p^2$.
This feature of the analytic structure of the propagator is
analogous to that found when studying classical wave propagation
in a random medium [10], as noted earlier.

\section {The Quark Self-Energy in The Kraichnan Random-Coupling
Approximation}

The approximation considered here is shown in Fig. 1. There the
cross-hatched region denotes the condensate and the wavy lines are
condensate gluons. When translated into momentum space, the
assumption of large correlation length $(L\rightarrow \infty)$
means that the condensate gluons may be taken to carry zero
momentum.

We write the quark self-energy as \be \Sigma(p)=A(p^2)\slr
p+B(p^2).\ee In the random-coupling approximation we have [See
Fig. 1], \be A(p^2)\slr p+B(p^2)=-\kappa^2\gamma_\mu\frac{1}{\slr
p-A(p^2)\slr p-B(p^2)-m_q^{cur}}\gamma^\mu,\ee where $m_q^{cur}$
is a ``current" quark mass and $\kappa^2\equiv g^2\phi_0^2/24$
contains a color factor of $4/3$.

Let us first consider the case $m_q^{cur}=0$. We see that Eq.
(4.2) exhibits a bifurcation phenomenon. There is a solution with
$B=0$; however, we will consider another solution with $B\neq 0$,
which breaks chiral symmetry. That solution is \be A(p^2)&=&-1\,,
\\\nonumber B(p^2)&=&2\sqrt{p^2+\kappa^2},
\,\,\,\,\,\,\,\,\,\,\,\,\,\,\,\mbox{for}\,\,p^2\,>\,-\kappa^2.\ee
and \be A(p^2)&=&\frac{1}{2}\left[1-\sqrt{1-8\kappa^2/p^2}\right],
\\\nonumber B(p^2)&=&0, \,\,\,\,\,\,\,\,\,\,\,\,\mbox{for}\,\,p^2\,<\,-\kappa^2.\ee It is
useful to define a dynamical mass \be
M(p^2)=\frac{B(p^2)+m_q^{cur}}{1-A(p^2)},\ee which for
$m_q^{cur}=0$ is \be
M(p^2)=(p^2+\kappa^2)^{1/2}\theta(p^2+\kappa^2).\ee It may be seen
that Eqs. (4.3)-(4.4) represent the $m_q^{cur}=0$ limit of a
continuous solution of Eq. (4.2) obtained with $m_q^{cur}\neq 0$.
Equation (4.2) may be solved for $m_q^{cur}\neq 0$. One finds that
$M(p^2)$ satisfies a fourth-order polynomial equation. That
equation has four solutions for each $p^2$. We choose a real
continuous solution such that $M(p^2)\rightarrow m_q^{cur}$ as
$p^2 \rightarrow -\infty$. The value of $M^2(p^2)$, $B(p^2)$ and
$[1-A(p^2)]$ for such solutions are shown in Figs. 2 to 14, for
various values of $m_q^{cur}$. The values chosen are appropriate
for up, down, strange, charm and bottom quarks. [We have also
taken $\kappa^2$=(232 Mev)$^2$.]

We note that for the solutions chosen, for large spacelike $p^2$
we have $A(p^2)\rightarrow 0, B(p^2)\rightarrow 0$ and \be <S(p)>
\simeq \frac{1}{\slr p-m_q^{cur}}.\ee For timelike $p^2$ we have
$M(p^2)\rightarrow p^2+\kappa^2$ for large $p^2$. It may be seen
that the equation $p^2=M^2(p^2)$ has no solution and therefore
there are no poles in the statistically averaged quark propagator
for real $p^2$. This generalizes the corresponding result of the
classical theory to the relativistic theory.

\begin{figure}
\includegraphics[bb=140 200 200 435, angle=0, scale=1]{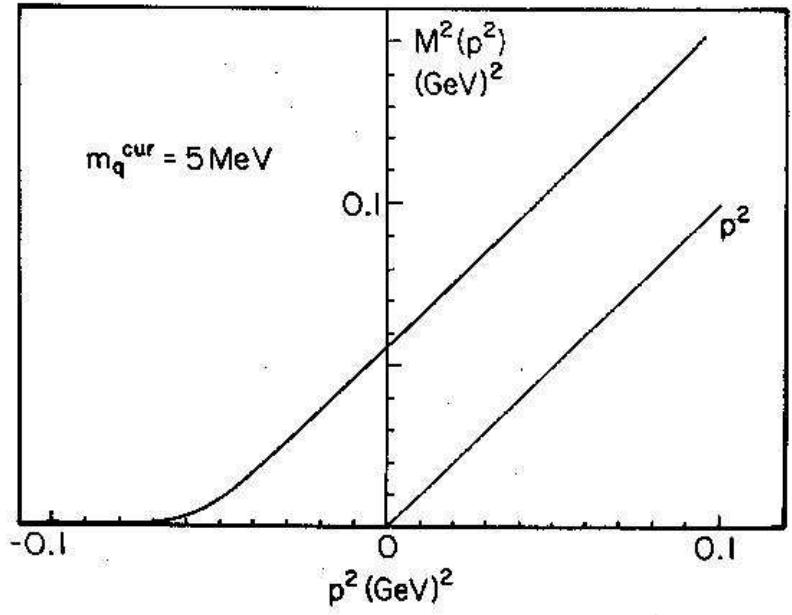}%
\caption{The square of the dynamical mass, $M^2(p^2)$, for up and
down quarks. Here we chose $m_q^{cur}$=5 MeV. [For $p^2 > 0$,
$M^2(p^2)=p^2+\kappa^2$.]}
\end{figure}

\begin{figure}
\includegraphics[bb=140 200 200 455, angle=0, scale=1]{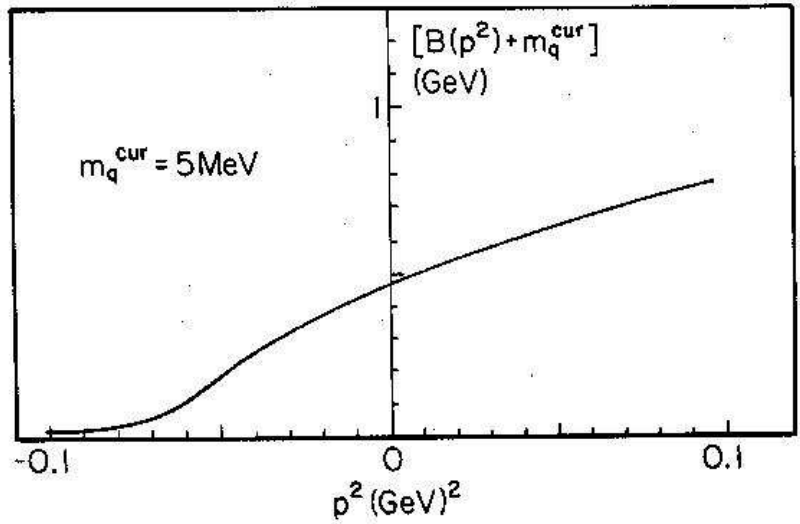}%
\caption{The mass parameter $[B(p^2)+m_q^{cur}]$ for up and down
quarks with $m_q^{cur}$ =5 MeV.}
\end{figure}

\begin{figure}
\includegraphics[bb=100 170 280 435, angle=0, scale=1]{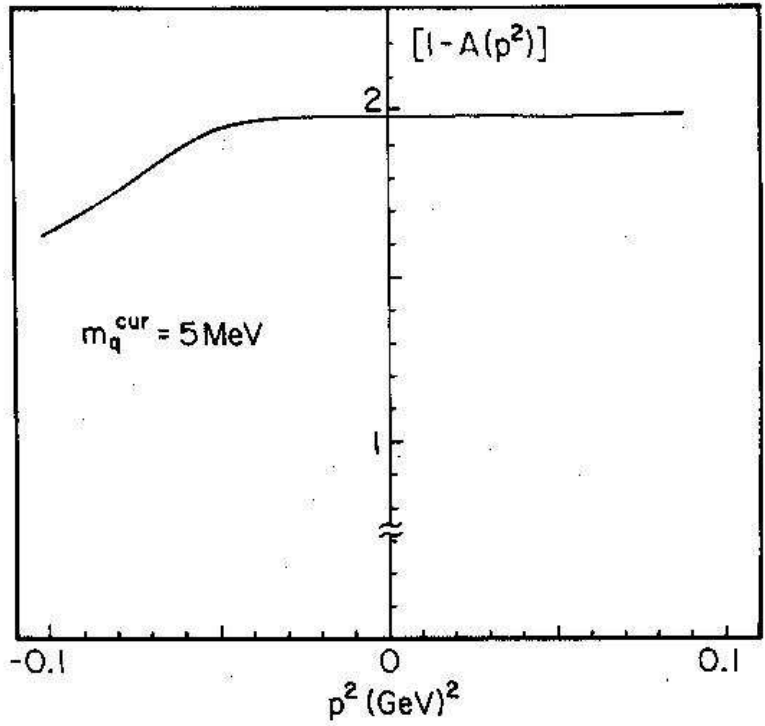}%
\caption{The wave function renormalization parameter $[1-A(p^2)]$
for up and down quarks.}
\end{figure}

\begin{figure}
\includegraphics[bb=100 200 280 435, angle=0, scale=1]{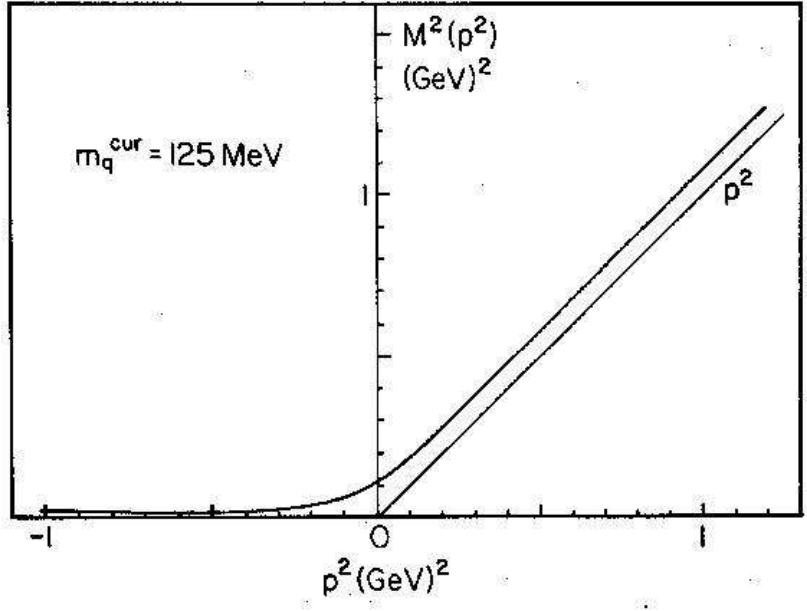}%
\caption{The square of the dynamical mass, $M^2(p^2)$, for the
strange quark. Here $m_q^{cur}$=125 MeV. (Note the change in scale
with respect to Figs. 2-4.)}
\end{figure}

\begin{figure}
\includegraphics[bb=100 200 280 435, angle=0, scale=1]{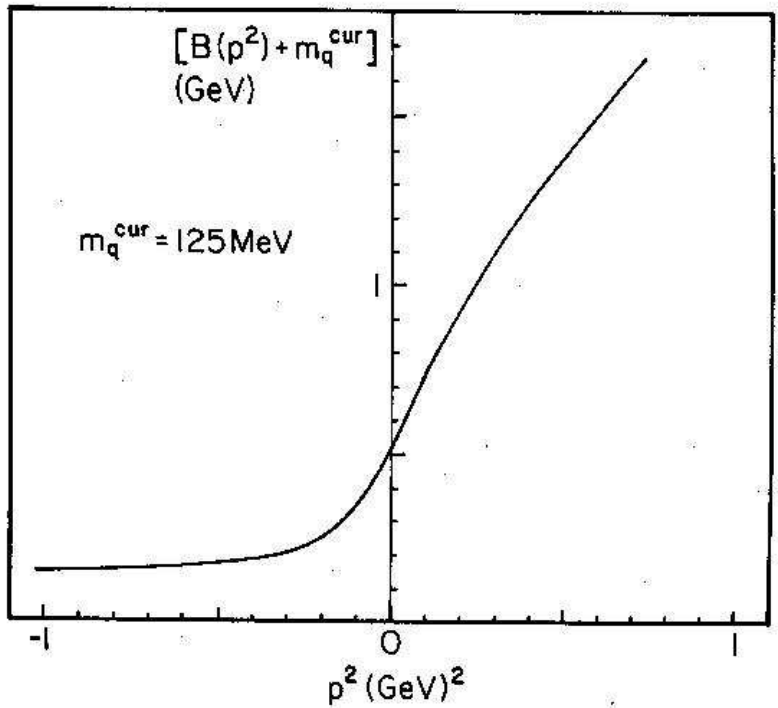}%
\caption{The mass parameter $[B(p^2)+m_q^{cur}]$ for the strange
quark. (Here $m_q^{cur}$  = 125 MeV.)}
\end{figure}

\begin{figure}
\includegraphics[bb=100 200 280 435, angle=0, scale=1]{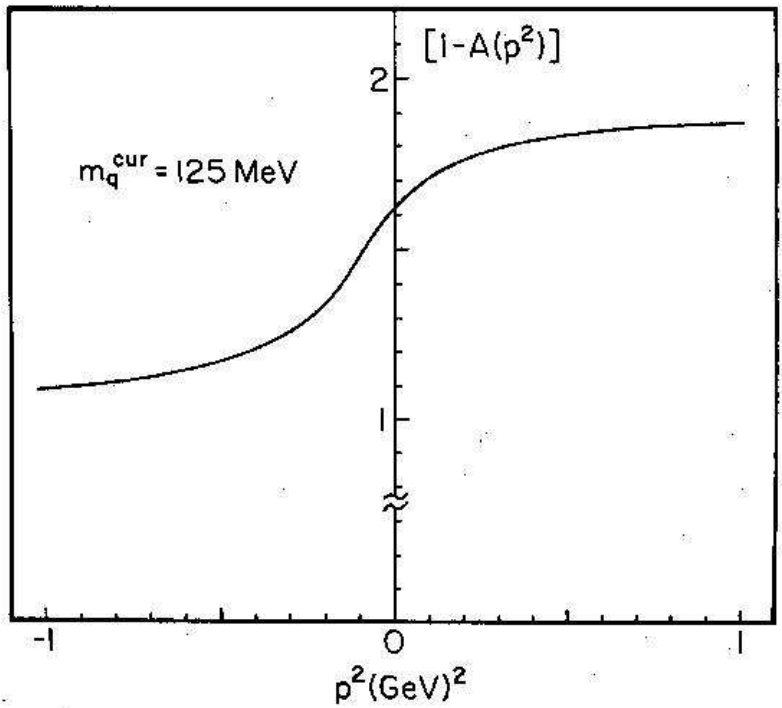}%
\caption{The wave function renormalization parameter $[1-A(p^2)]$
for the strange quark. [See caption to Fig. 5.]}
\end{figure}

\begin{figure}
\includegraphics[bb=100 200 280 435, angle=0, scale=1]{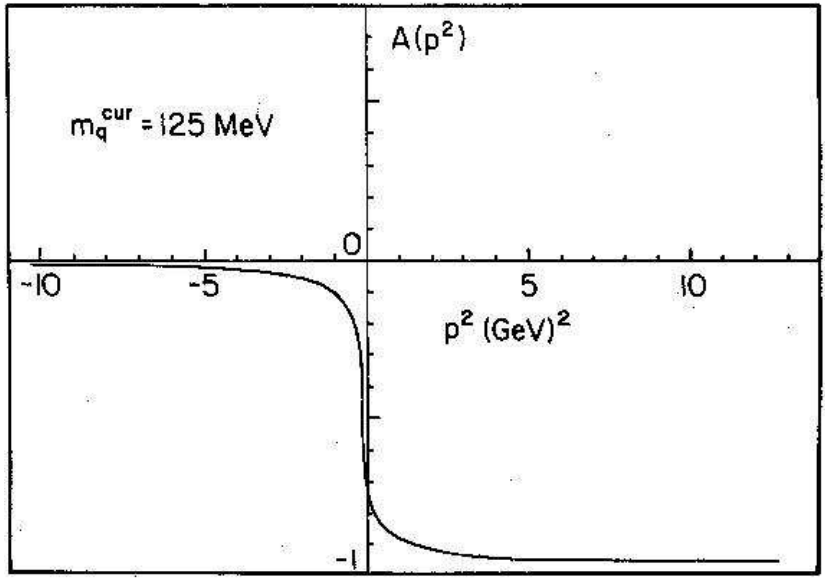}%
\caption{The function $A(p^2)$ for the strange quark
$(m_q^{cur}$=125 MeV) for an extended range of values of $p^2$.}
\end{figure}

\begin{figure}
\includegraphics[bb=100 210 280 435, angle=0, scale=1]{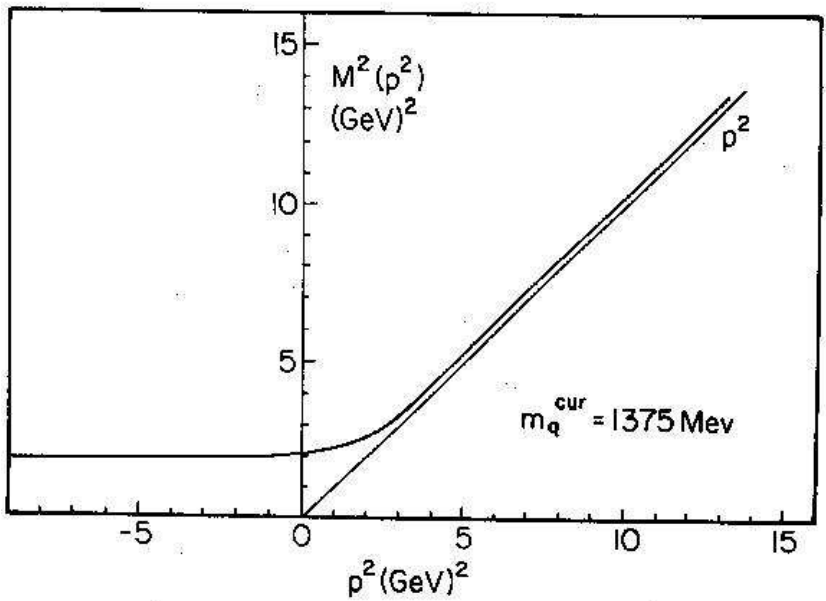}%
\caption{The square of the dynamical mass, $M^2(p^2)$, for the
charm quark. Here $m_q^{cur}$ = 1375 GeV. (Note the change of
scale with respect to Figs. 2-8.)}
\end{figure}

\begin{figure}
\includegraphics[bb=100 220 280 465, angle=0, scale=1]{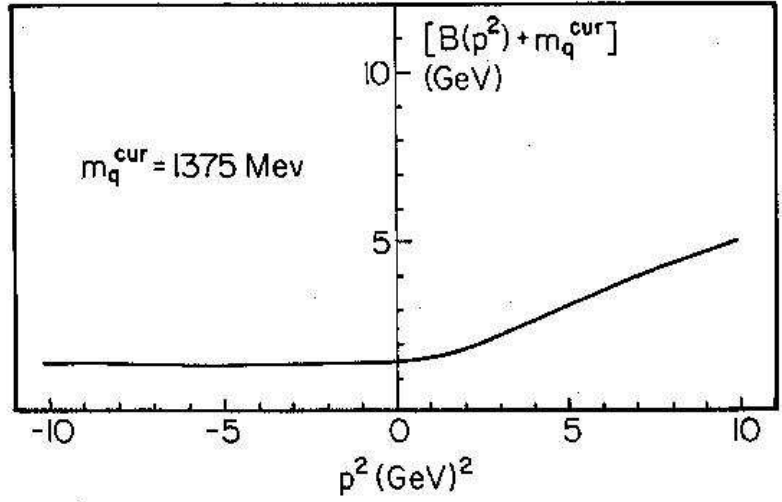}%
\caption{The mass parameter $[B(p^2)+m_q^{cur}]$ for the charm
quark. (Here $m_q^{cur}$ = 1375 MeV.) [See caption to Fig. 9.]}
\end{figure}

\begin{figure}
\includegraphics[bb=0 200 300 435, angle=0, scale=1]{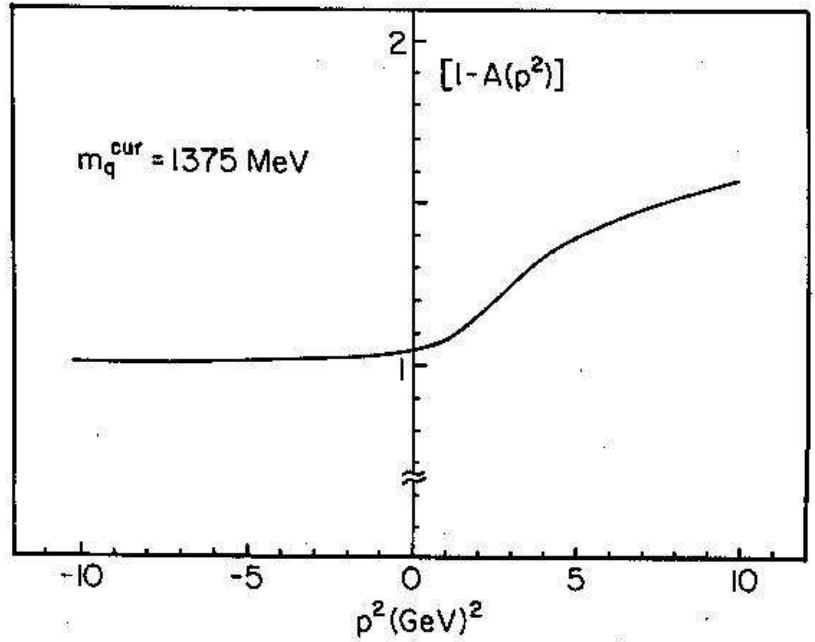}%
\caption{The wave function renormalization parameter, $[1-A(p^2)]$
for the charm quark. Here $m_q^{cur}$ = 1375 GeV. (Note the change
of scale with respect to Figs. 2-8.)}
\end{figure}

\begin{figure}
\includegraphics[bb=100 240 300 435, angle=0, scale=1]{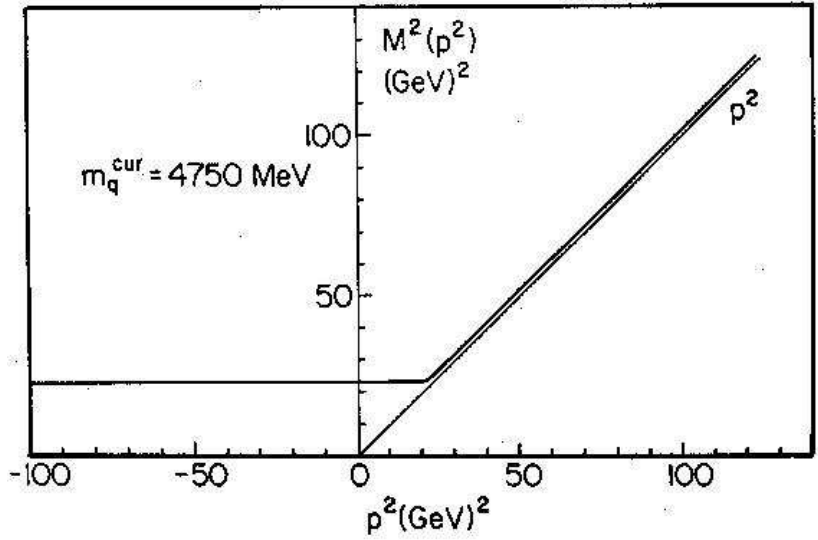}%
\caption{The square of the dynamical mass, $M^2(p^2)$, for the
bottom quark. Here $m_q^{cur}$ = 4750 MeV. (Note the change of
scale with respect to Figs. 2-11.)}
\end{figure}

\begin{figure}
\includegraphics[bb=100 200 300 455, angle=0, scale=1]{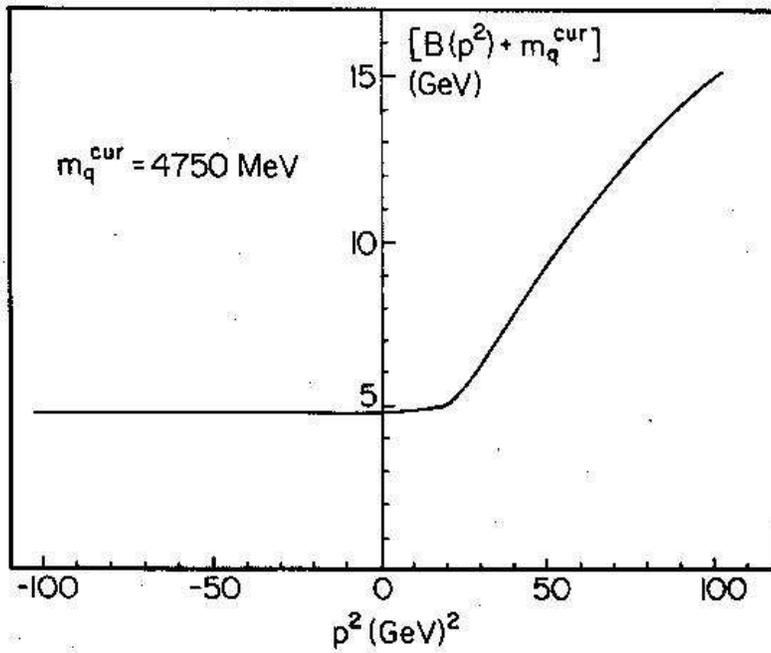}%
\caption{The mass parameter $[B(p^2)+m_q^{cur}]$ for the bottom
quark. Here $m_q^{cur}$ = 4750 MeV. [See caption to Fig. 12.]}
\end{figure}

\begin{figure}
\includegraphics[bb=100 200 300 435, angle=0, scale=1]{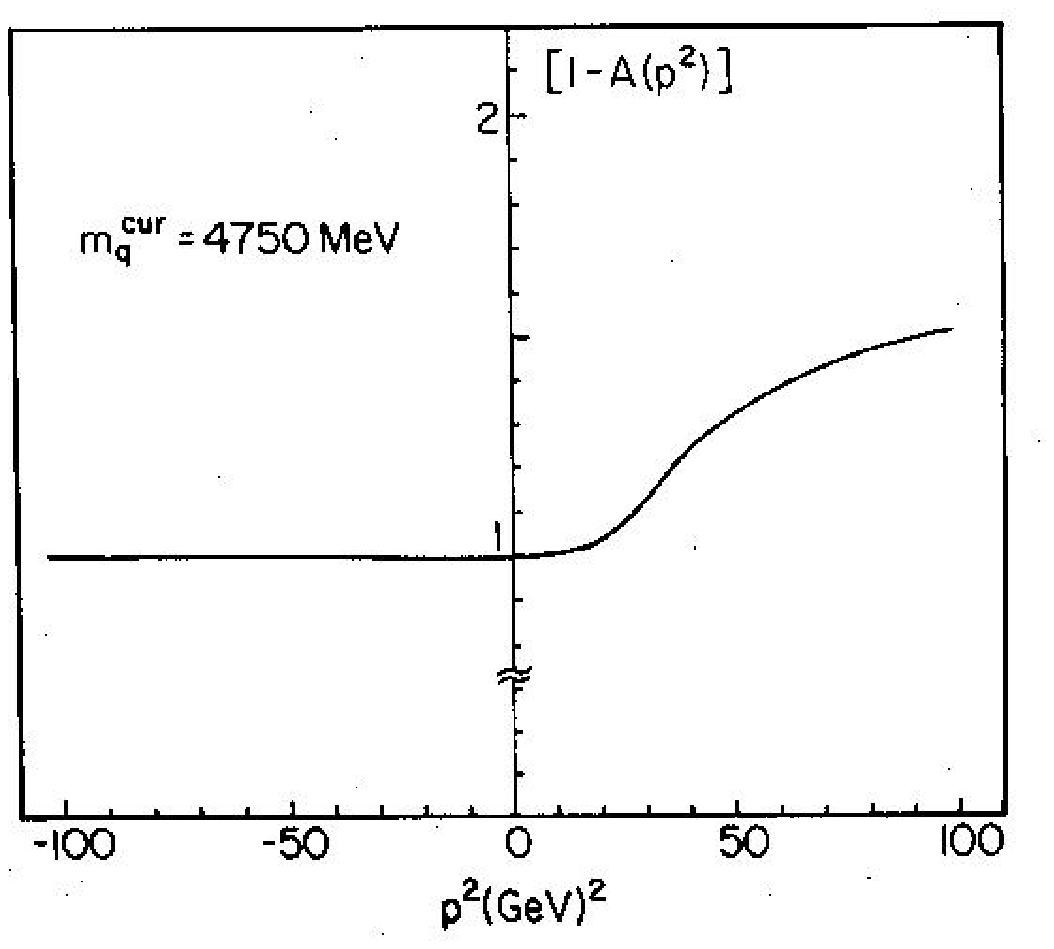}%
\caption{The wave function renormalization parameter $[1-A(p^2)]$
for the bottom quark. Here $m_q^{cur}$ = 4750 MeV. [See caption to
Fig. 12.]}
\end{figure}

\newpage
\section {Discussion}

In this work we have shown that the ensemble average of the quark
propagator exhibits damping. A similar result was obtained for the
gluon propagator in an earlier work [12] where we used the
first-order smoothing approximation [13] to generate an expression
for the vacuum polarization operator. In Ref. [12] we studied the
gluon propagator in momentum space and carried out a Fourier
transformation to coordinate space. In that case one could see
explicitly how the absence of on-mass-shell singularities in the
propagator leads to damping of the wave both for spacelike and
timelike propagation. In that analysis, carried in the Landau
gauge, it was seen that the gluon obtained a dynamical mass
$m_G^2=g^2\phi_0^2/4$. We also note that in a lattice simulation
of $SU_c$(3) Yang-Mills theory in the Landau gauge [14] it was
found the the gluon obtained a dynamical mass of $m_G=600\pm 90$
MeV. Since we  had put $\kappa^2=g^2\phi_0^2/24$, we use this
information to fix the value of $\kappa$ [$\kappa$=232 MeV].

In the work reported here we find that the solution which exhibits
nonpropagation of quarks over large distances is also
characterized as having $B\neq 0$. Nonzero values for $B$ denote
chiral symmetry breaking. It is seen that only a single parameter,
$g^2\phi_0^2$, governs both phenomena. This is a satisfactory
result in the case of QCD, since only a single dynamically
generated dimensionful parameter should characterize the theory.

$\\$ \textbf{Acknowledgment} This work was supported in part by a
grant from the Faculty Award Research Program of the City
University of New York awarded to Xiangdong Li.


\end{document}